\documentclass{article}
\usepackage{epsf,graphicx,amssymb,amsmath,esint}

\newcommand{\PTm}{{$\mathcal{PT}$}}

\newcommand{\be}{\begin{equation}}
\newcommand{\en}{\end{equation}}
\newcommand{\bea}{\begin{eqnarray}}
\newcommand{\ena}{\end{eqnarray}}
\newcommand{\ee}{{{\mathrm{e}}}}

\newcommand{\ii}{{{\mathrm{i}}}}

\newcommand{\abs}[1]{{\left|#1\right|}}
\newcommand{\odk}[1]{(\ref{#1})}

\newcommand{\ve}{\varepsilon}

\newcommand{\odkf}[1]{Fig.~\ref{#1}}
\newcommand{\ks}[1]{#1^*}

\newcommand{\re}{\mathop{\mathrm{Re}}}

\newcommand{\bes}{\begin{subequations}\bea}
\newcommand{\ens}{\ena\end{subequations}}

\begin{document}
\mark{{}{}}
\title{Spectra of \PTm-symmetric Hamiltonians on Tobogganic Contours}

\author{Hynek B\'ila}

\maketitle

\section{Introduction}

The term \PTm-symmetric quantum mechanics, although defined to be of a much broader use, was coined in tight connection with C. Bender's analysis of one-dimensional Schr\"odinger Hamiltonians with potentials 
\be
V(x)=x^2({\ii}x)^{\ve},\label{potdef}
\en
see \cite{PTSQM} (we will call them Bender Hamiltonians in the following). This class of operators -- characterised by one real parameter $\ve$ -- contained the probably most exploited Hamiltonian in the history of physics: the $\ve=0$ harmonic oscillator; but on the other hand, the other members of the family were strange Hamiltonians with imaginary potentials which do not appear physical at all. The aim of the suggested \PTm-symmetric treatment was to give them an acceptable interpretation and it became a rather standardised procedure since then. One of the key facts which allowed for the physical interpretation was that the spectrum of these operators were real at least for some non-zero values of $\ve$, a fact which stems from observation that first, for \PTm-symmetric operators the complex eigenvalues appear in mutually conjugated pairs, and second, that the spectrum depends continuously on $\ve$. Thus the boundary between real-energy and complex-energy domains can lie only at an \emph{exceptional point}, \emph{i.e.} a point where at least two eigenvalues merge. Consequently one has to depart at least some finite distance on the $\ve$-axis from the non-degenerate Harmonic oscillator to see any selected eigenvalue complexify. What actually happens in the discussed case is visible at the classical figure \odkf{nontob}: the spectrum is real for all positive $\ve$, while at negative $\ve$ the lower the energy is the longer it remains real. 

\begin{figure}
\includegraphics[width=\textwidth]{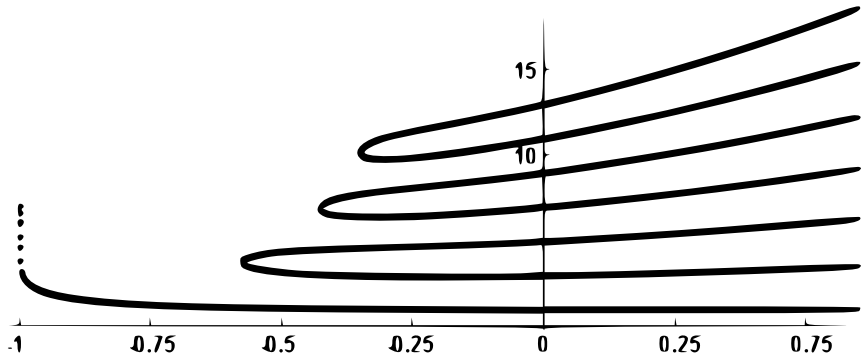} \label{nontob}
\caption{Well-known dependence of the eigenvalues of the Bender Hamiltonian on $\ve$, non-tobogganic case. Complex energies are not shown. Based on numerical calculation it is expected that the value $\ve_n$ at which the $n$-th energy becomes complex, tends to zero as $n\rightarrow\infty$.}
\end{figure}

However, the validity of the continuous energy dependence assumption is not obvious. An immediate question of the reader could be: What happens when $\ve=2$, which yields an obviously unphysical $-x^4$ potential with unbounded continuous spectrum? It is well known now, that the correct response has to deal with boundary condition in the complex plane. The Hamiltonian which is \emph{the proper} continuation of the Harmonic oscillator in the parameter $\ve$ is defined on space of functions which are square-integrable not on the real line, but on some asymptotically straight contour in the complex plane which lies in the correct \textit{Stokes sector} of the Schr\"odinger equation.  Here, correct means that the sectors are itself a continuation of those sectors which contain the real line in the $\ve=0$ case. These sectors (also called ``wedges'') are turning down in the complex plane as $\ve$ increases and for $\ve\geq2$ they no more contain the real line. Therefore, the conventional $-x^4$ Hamiltonian on $L_2(\mathbb{R})$ is not the only Hamiltonian which deserves this name; in some sense its analogue defined by the same differential equation but with complex boundary conditions is more natural. 

For sake of clarity let us recall that for potentials we are dealing with the exact choice of the integration contour is irrelevant as long as it lies inside the Stokes sector asymptotically. It became customary to omit mentioning the contour altogether, speaking only about boundary conditions imposed inside a particular sector. On the other hand, the contour is a convenient means for defining the scalar product and, practically, some concrete choice of the contour is necessary for numerical computation; thus we will speak about integration contours rather than wedge-defined boundary conditions in the rest of the paper, keeping in mind that due to the potential's analyticity distinct contours with same-wedge asymptotics are equivalent\footnote{In fact one can even release the asymptotic straightness condition provided that the contour does not oscillate too rapidly in the asymptotic region. The technical details of equivalence between contour integrability and boundary conditions in infinity are beyond the scope of this article.}. It may be also noted that in the special case $\ve=2$ it is enough to pose the boundary conditions on the shifted real line, \emph{i.e.} $x=t-\ii c$ with arbitrary positive constant $c$ ($t$ is a real parameter). For higher $\ve$ one has to use bent contours to remain inside the Stokes wedge.

The existence of aforementioned continuation is interesting from different points of view. For example, it is relevant to the Dyson argument about convergence of the perturbative series, which is roughly stated as follows: having a potential $x^2+gx^4$ (or its field-theoretical counterpart, the argument was originally formulated for quantum electrodynamics \cite{dys}) one may use perturbative expansion in $g$, but since for any negative $g$ the spectrum collapses to $(-\infty,\infty)$ and the perturbative calculation cannot produce this, the convergence radius of the series ought to be zero and the series is at best asymptotic. The existence of a negative-$g$ Hamiltonian with discrete real and below bounded spectrum invalidates the argument's core assumption: it is now entirely possible that the series converges and gives the spectrum of the complex-plane $-x^4$ potential when evaluated at negative $g$, \emph{viz.} \cite{MAHBH}.

\section{Quantum Toboggans}

The observation which we want to point out now is that to force the integration contour to be asymptotically straight and inside the correct Stokes wedges is insufficient to uniquely define the spectra of Hamiltonians with \odk{potdef}. One has to take into account that for non-integer $\ve$ the potential lives on multiple Riemann sheets and it matters how the integration contour is distributed upon the sheets. In our presently discussed class of Hamiltonians, the only singularity lies at $x=0$, which allows us to use only one \emph{winding number} $\lambda$ to fully characterise the contour. Hence instead of one \PTm-symmteric continuation of the harmonic oscillator we obtain an infinite series numbered by distinct integer values of $\lambda$. The $\lambda=0$ trajectory represents the only usually considered case. The higher-$\lambda$ contours define distinct Hamiltonians which are sometimes referred to as \emph{quantum toboggans}, the reason of such denomination is clear when one imagines the Riemann sheets forming a helicoid\footnote{Suggested imagination is, of course, consistent only when infinite nuber of sheets are present.}.

A technical note: To take profit from the fact that real eigenvalues have to merge before becoming complex, one must take care about the \PTm-symmetry. In particular it dictates the position of the branch cut. It has to be chosen to lead upwards along the positive imaginary axis. On the other hand, the potential has to be \PTm-symmetric on the contour, which, for $\lambda=0$, forces the contour to pass below the singularity. If one wanted to have the contour bypassing zero from above, which is an equally reasonable option for the harmonic oscillator where no singularity is present, one has to change the potential accordingly to preserve the \PTm-symmetry of the potential \emph{on the contour}. Without much surprise this yields
\be
V=x^2(-{\ii} x)^\ve \label{minuspot}
\en
Such change of potential together with the change of the contour would obviously be nothing than the reparametrisation $x\rightarrow\ks{x}$; as such it leaves the spectrum intact. We can similarly disregard the differences within analogous mutually conjugated pairs of trajectories and potentials even in more complicated settings. Therefore, without loss of generality, we will stick to the definition \odk{potdef} and keep in mind that also the contour must conform to the \PTm-symmetry.

\begin{figure}
\includegraphics[width=\textwidth]{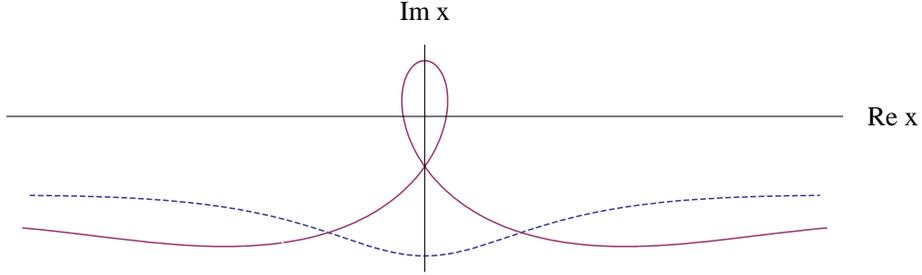} \label{kontury}
\caption{Considered integration contours: a straight one (winding number $\lambda=0$) is drawn in the dashed line, while its most elementary tobogganic counterpart ($\lambda=1$) is depicted in solid line. The latter contour is consistent with \PTm-symmetry only if the the branch cut aims downwards, therefore we need to use an inverted one with potential \odk{potdef}.}
\end{figure}

The integration contours are schematically drawn in \odkf{kontury} and \odkf{kontury2}. It turns out that the choice of the $\lambda=1$ contour in \odkf{kontury} is inconsistent with \PTm-symmetry, instead one has to use the up-down inverted curve, as depicted in \odkf{kontury2}. A general rule is that the center point of the contour lies below zero (for concreteness say at $x=-\ii$) on the principal sheet, which is the one where $V(-\ii)$ is real.

\begin{figure}
\includegraphics[width=\textwidth]{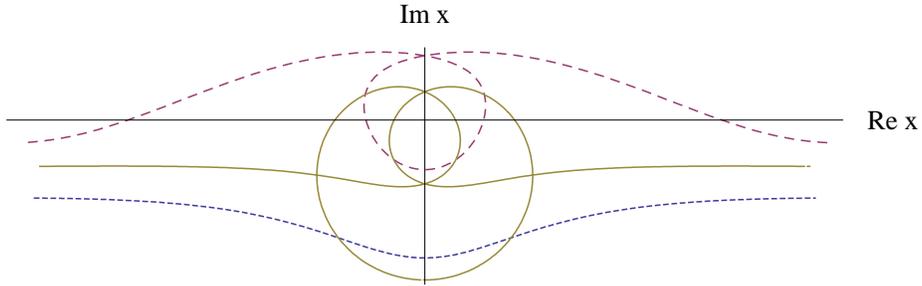} \label{kontury2}
\caption{\PTm-symmetric contours for $\lambda=1,2,3$ of the potential $x^2(\ii x)^\ve$. In contrast with \odkf{kontury} the $\lambda=1$ contour is chosen inverted with respect to the real axis, conforming the \PTm-symmetry assumption.}
\end{figure}

\section{Numerical Analysis}

Unfortunately, the considered potentials are not exactly solvable. Therefore we have to rely solely on numerical treatment. We use a relatively straightforward method for computation of eigenvalues; we are interested whether real eigenvalues are present -- the existence of complex part of the spectrum can be partially deduced from the distribution of the exceptional points -- this allows us to significanly simplify the calculation. First of all, a starting energy is chosen. Then we calculate two independent solutions of the differential equation with the chosen energy; let us denote them $\psi_1$ and $\psi_2$. The equation has been solved on the contour parametrised as
\bes
x & = & t-\ii\quad \textrm{for}\ \lambda=0,\\
x & = & \ii\ee^{\ii t}\Theta(\pi-\abs{t})+(t-\pi-\ii)\Theta(\abs{t}-\pi)\quad \textrm{for}\ \lambda=1,\\
x & = & -\ii\ee^{\ii t}\Theta(2\pi-\abs{t})+(t-2\pi-\ii)\Theta(\abs{t}-2\pi)\quad \textrm{for}\ \lambda=2,
\ens
and so on; \textit{i.e.} the contour consists of a circle of unit radius $\lambda$-times encircling the singularity in $x=0$  and a straight line parallel with the real axis\footnote{This choice puts a limit on applicability of the method to $\ve<2$.} matched to the circle at $x=-\ii$. For concreteness the initial conditions for $\psi_{1,2}$ are set in the centre of the contour (i.e. $t=0$) to satisfy
\bes
&\psi_1(0)=0,\ \psi_1'(0)=1,& \label{urug1}\\
&\psi_2(0)=1,\ \psi_2'(0)=0.& \label{urug2}
\ens

The number $E$ is an eigenvalue if there exists a linear combination of $\psi_1$ and $\psi_2$ which is integrable. Because the asymptotics of the solution is exponential, this is equivalent with the existence of a linear combination tending to zero. In our calculations it is satisfactory to look at the function values at $x_\pm\approx\pm10$, at least for the lowest eigenstates. The \PTm-symmetry of the potential plays now a key r\^ole. Since we are interested only in the real part of the spectrum we can assume the \PTm-symmetry of the wave function. The equation for the eigenvalues, which originally reads\footnote{The energy dependence of the solution is made explicit in the following.}
\be
\det \left( 
\begin{array}{cc}
\psi_1(E,x_+) & \psi_1(E,x_-) \\
\psi_2(E,x_+) & \psi_2(E,x_-)
\end{array} 
\right) =0
\label{determrov}
\en
is now, having in mind \odk{urug1} and \odk{urug2}, simplified into 
\be
\re \ks{\psi_1(E,x_+)}\psi_2(E,x_+)=0. \label{rerov}
\en
Note that the left hand side of \odk{determrov} is a complex function of energy, whereas the left hand side of \odk{rerov} is real. This is clearly an advantage -- the zeros of a real function can be determined by the bisection method. The described procedure works well for the case $\ve=0$ where comparison with the exact results is feasible; the exact result is recovered up to five- or six-digits precision, the precision can be enhanced with some loss of speed. Comparison of the non-tobogganic case with results obtained in \cite{PTSQM} shows no significant differences. When $\ve\rightarrow 2$ the computation becomes slower if the precision of computation has to be maintained since the solution's asymptotics becomes more vulnerable to numerical errors.

\begin{figure}
\includegraphics[width=0.45\textwidth]{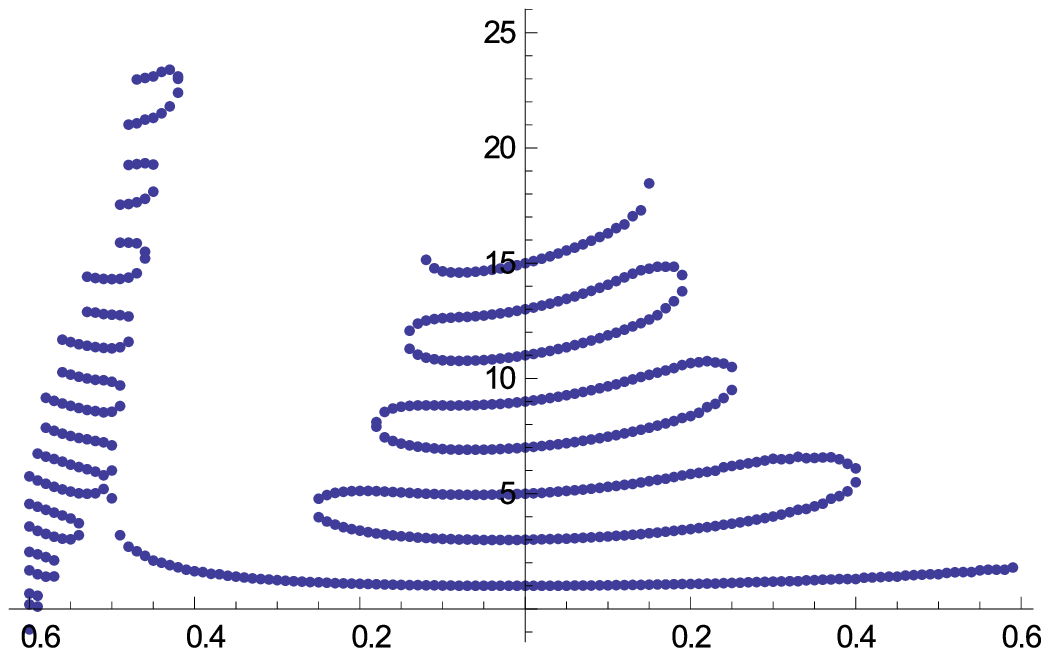}
\includegraphics[width=0.45\textwidth]{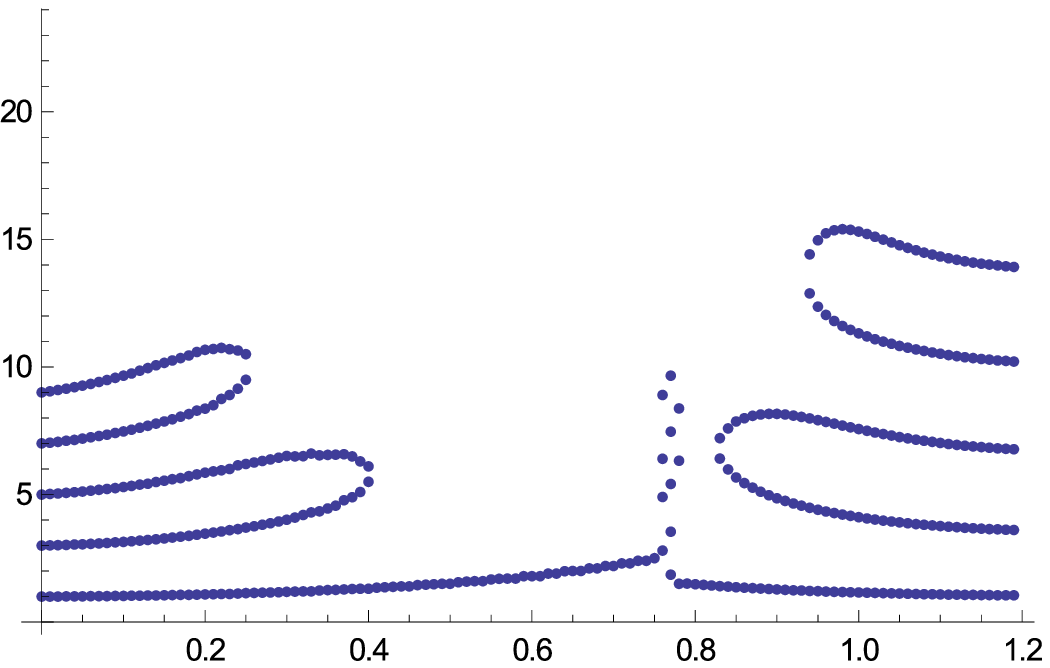}
\label{tobog2}
\caption{Spectra of Bender Hamiltonian, contour with $\lambda=1$. The graph on the right hand side illustrates the behaviour around $\ve=1$ where real eigenvalues emerge again. The algorithm was requested to find only lowest six real eigenvalues (five on the right graph), hence the lines corresponding to higher eigenvalues are discontinued where lower eigenvalue pair emerges. The chaotically distributed points in the upper part of the graph result from the error in the algorithm's implementation which occurs when the demanded number of real eigenvalues is not found. On both graphs, the vertical axis represents the energy while $\ve$ is drawn horizontally.}
\end{figure}

\begin{figure}
\includegraphics[width=0.45\textwidth]{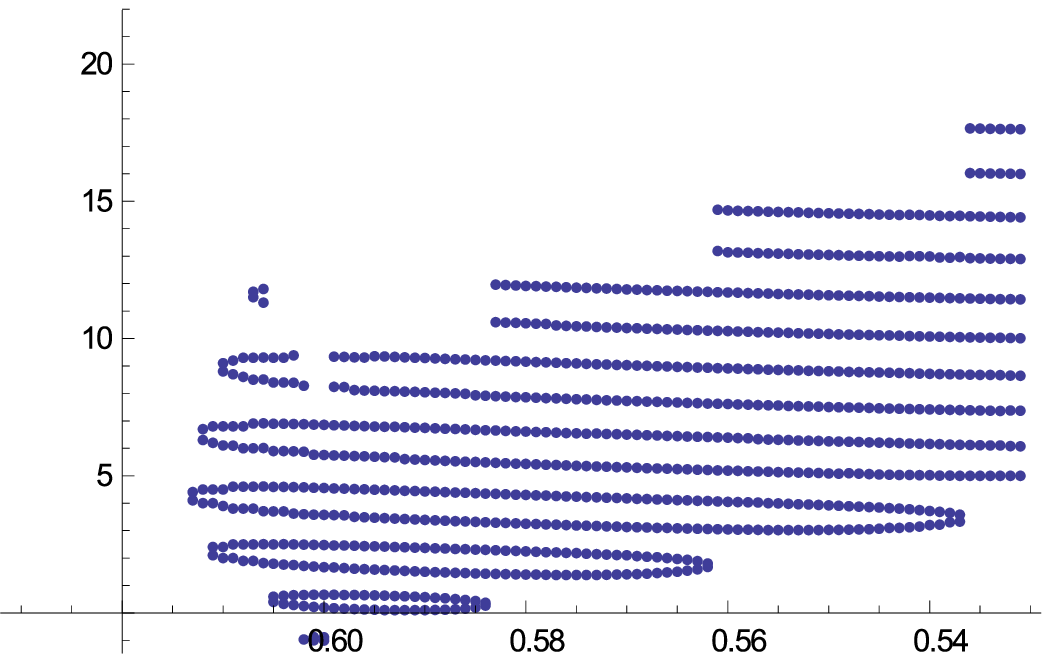}
\includegraphics[width=0.45\textwidth]{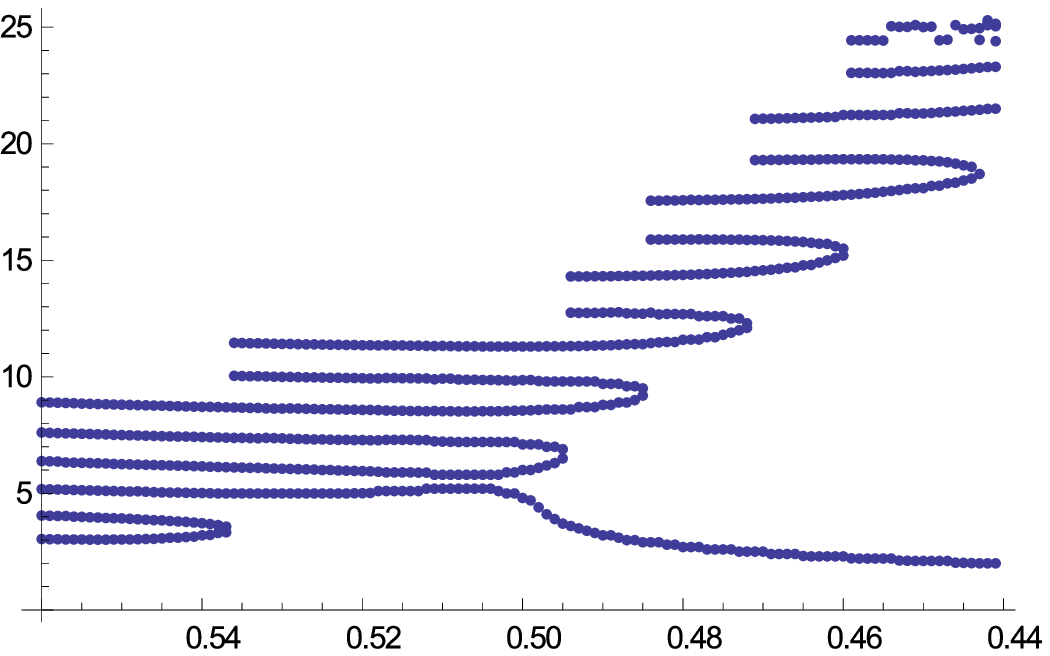}
\caption{A detail of the critical region between $\ve=-0.45$ and $\ve=-0.65$ for $\lambda=1$.}
\label{tobog7}
\end{figure}

The results show that the behaviour of eigenvalues depends on the winding number $\lambda$. The results for $\lambda=1$ (see \odkf{tobog2}) exhibit vast differences from the $\lambda=0$ case. First, except the lowest eigenvalue the spectrum complexifies also at $\ve>0$. As \odkf{tobog2} suggests, the region of reality is broader for the low lying eigenvalues. It is entirely possible that there does not exist either left or right neighborhood of $\ve=0$ where the whole spectrum is purely real, in contradistinction to the non-tobogganic contour which yields real spectra for all $\ve>0$. However it seems that there is a previously unattested region of real eigenvalues at $\ve<-0.4$, probably not perturbatively accesible since perturbative calculations usually break down in exceptional points. The lowest energy tends to infinity as $\ve\rightarrow-1$ in $\lambda=0$ case; if $\lambda=1$ it joins the other real eigenvalues in the left region and eventually complexifies in an exceptional point near $\ve=-0.61$ (see \odkf{tobog7}) -- it is interesting to note that here it does not represent the ground state. The spectrum has to also be real in the vicinity of $\ve\in \mathbb{N}$ since the singularity disappears there and the contours for different $\lambda$ are equivalent, consequently the real spectrum of $\lambda=0$ case must be reproduced (see \odkf{tobog2}). 

\begin{figure}
\includegraphics[width=0.45\textwidth]{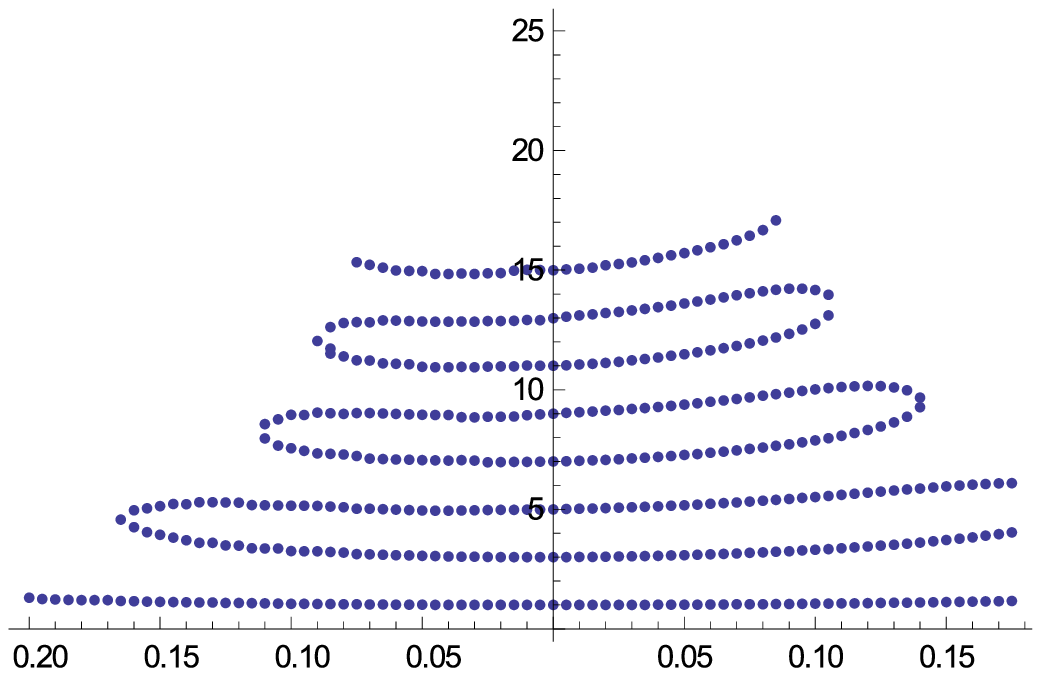}
\includegraphics[width=0.45\textwidth]{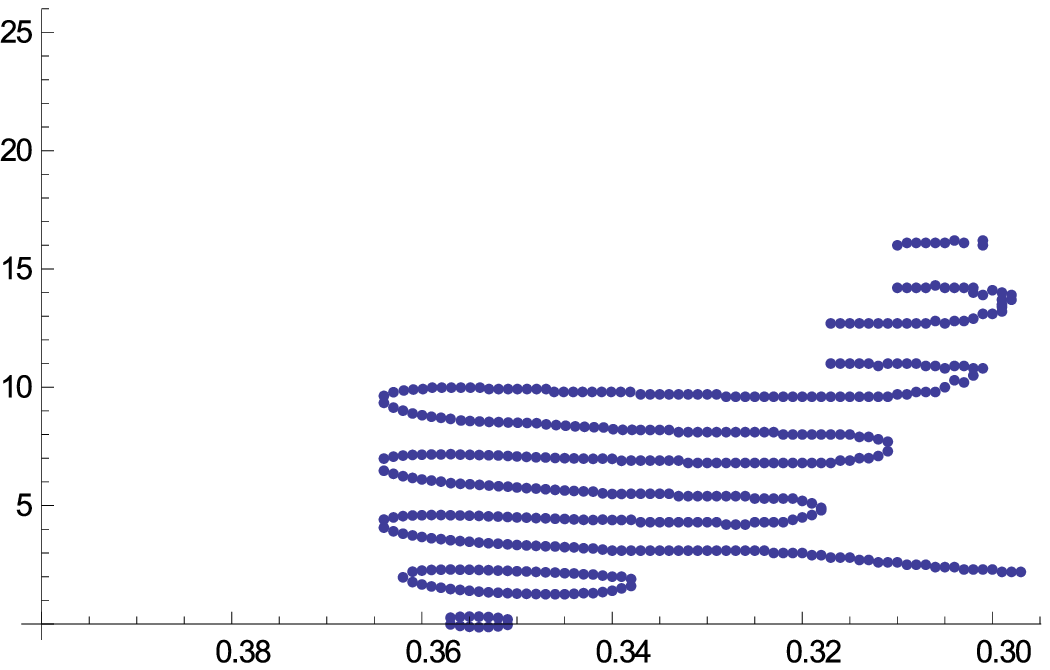}
\label{tobog3}
\caption{The $\lambda=2$ case. The region of reality for each eigenvalue is narrower than for $\lambda=1$. On the right hand side the critical region is again depicted in a greater detail, its position is moved further to the right with respect to $\lambda=1$ case.}
\end{figure}

The overall picture does not change significantly when $\lambda=2$. The overall pattern is similar to $\lambda=1$ (see \odkf{tobog3}). Possibly another region of real spectrum exists near $\ve=-1$, but the computation becomes lengthy and unreliable in those points, since near $\ve=-1$ the solutions do not decrease enough rapidly; we are not confident in results obtained in this region by the above described method and leave this problem for future investigation.

\section{Summary}

The introduction of tobogganic contours into the Bender potentials produces another versions of \PTm-symmetric Hamiltonian. Though they are closely related to the original non-tobogganic Hamiltonian, they are indeed different and exhibit qualitatively distinct behaviour with exceptional points standing between intervals of reality including the points $\ve=n, n\in\mathbb{N}$. In an alternative approach, one can change the variable to unbend the contour, this leaves e.g. the $\lambda=1$ Hamiltonian in the form
\be
-\frac{\psi''}{9y^4}-\frac{2\psi'}{y}+\ii^\ve y^{6+3\ve}.
\en
after putting $x=y^3$. Such transformations were discussed in \cite{tobogz}. They are interesting as a mehod that allows to simply transform the problem to an ordinary differential equation of one real variable. On the other hand the Schr\"odinger-like form of the Hamiltonian is lost, which makes the example less physcally appealing.

If we are interested only in the transition between the harmonic oscillator and the negative quartic oscillator, the choice of $\lambda$ is clearly irrelevant. For integer $\ve$ there is no singularity and distinctly winded contours must yield identical spectra. Therefore it can be said that any tobogganic Hamiltonian defines good continuation of the harmonic oscillator, and the $\lambda=0$ special case is only ``incidentally'' privileged due to its real spectrum.

\begin{figure}
\includegraphics[width=0.45\textwidth]{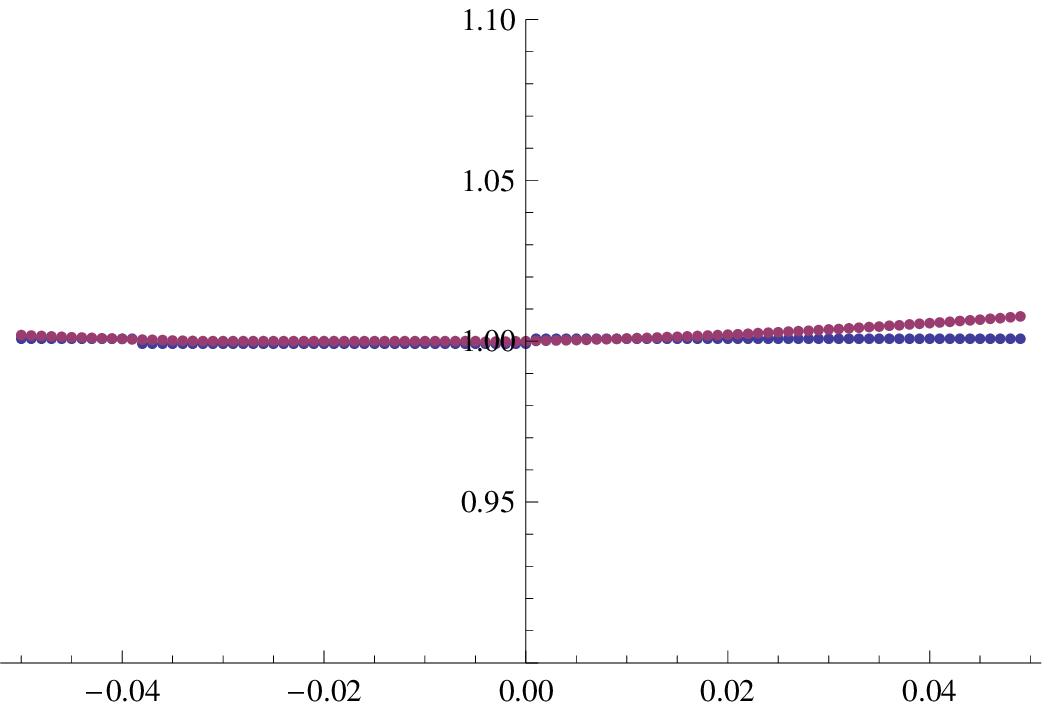}
\includegraphics[width=0.45\textwidth]{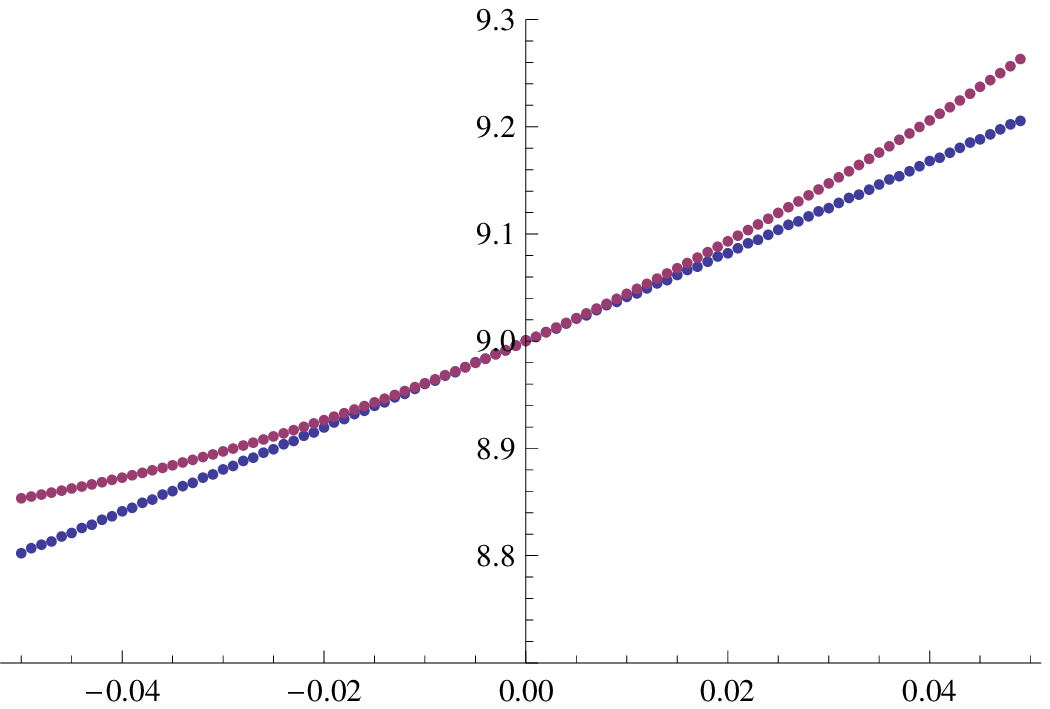}
\caption{The first (left) and fifth (right) eigenvalue plotted against $\ve$ for $\lambda=0$ and $\lambda=1$ (the latter is the less straight dependence).}
\label{tobogsrv}
\end{figure}

It may be also noted that the dependence on $\lambda$ is non-perturbative in $\ve$. The equality of the linear approximation coefficients for distinct $\lambda$ is visible from \odkf{tobogsrv}. Up to the first order the energy is, independently of the contour selection,
\be
E_n=2n+1+\frac{\ve}{2}\,\digamma\!\left(\frac{2\lceil n/2\rceil+1}{2}\right),\label{lunochod}
\en
$\digamma$ is the digamma function\footnote{$\digamma(x)=(\log\Gamma(x))'$ where $\Gamma$ is the standard Euler Gamma function.}. 

\subsection*{Acknowledgement:} This work was supported by the Czech Ministry of Education, Youth and Sports (Project LC06002).

\newcommand{\clj}[7]{\bibitem{#1} #2, \textit{#3}, #4 \textbf{#5} (#6) #7}
\newcommand{\cljo}[6]{\\ #1, \textit{#2}, #3 \textbf{#4} (#5) #6}
\newcommand{\cla}[4]{\bibitem{#1} #2, \textit{#3}, \texttt{arXiv:#4}}
\newcommand{\clao}[3]{\\ #1, \textit{#2}, \texttt{arXiv:#3}}
\newcommand{\kni}[4]{\bibitem{#1} #2, \textsl{#3}, #4}

\end{document}